\documentclass[letter, traditabstract]{aa} 

\usepackage{graphicx}
\usepackage{lscape}
\usepackage{soul}
\usepackage[varg]{txfonts}
\usepackage{upgreek} 
\usepackage{bigdelim}
\usepackage{multicol}
\usepackage{multirow} 
\usepackage[FIGTOPCAP, center, nooneline]{subfigure}
\usepackage{natbib}
\usepackage{rotating}
\usepackage{lscape}
\usepackage{float}
\usepackage{color}
\usepackage{amsmath}

\begin{document} 

\title{A study of C$_4$H$_3$N isomers in TMC-1: line by line detection of HCCCH$_2$CN\thanks{Based on observations with the 40-m radio telescope of the National Geographic Institute of Spain (IGN) at Yebes Observatory (projects 19A003 and 20A014). Yebes Observatory thanks the ERC for 
funding support under grant ERC-2013-Syg-610256-NANOCOSMOS.}
}

\author{
N. Marcelino\inst{\ref{inst1}}
\and
B. Tercero\inst{\ref{inst2},\ref{inst3}}
\and
M. Ag\'undez\inst{\ref{inst1}}
\and
J. Cernicharo\inst{\ref{inst1}}
          }

\institute{
Grupo de Astrof\'isica Molecular, Instituto de F\'isica Fundamental, CSIC, C/ Serrano 123, 28006 Madrid, Spain\label{inst1} \\
\email{nuria.marcelino@csic.es}
\and
Observatorio Astron\'omico Nacional (IGN), C/ Alfonso XII 3, 28014 Madrid, Spain\label{inst2}
\and
Observatorio de Yebes (IGN). Cerro de la Palera s/n, 19141 Yebes, Guadalajara, Spain\label{inst3}
}

\date{Received ; accepted }


\abstract
{We present Yebes 40m telescope observations of the three most stable C$_4$H$_3$N isomers 
towards the cyanopolyyne peak of TMC-1. We have detected 13 transitions from CH$_3$C$_3$N 
(A and E species), 16 lines from CH$_2$CCHCN, and 27 lines ($a$-type and $b$-type) from 
HCCCH$_2$CN. We thus provide a robust confirmation of the detection of HCCCH$_2$CN and CH$_2$CCHCN in space. 
We have constructed rotational diagrams for the three species, and obtained rotational 
temperatures between $4-8$\,K and similar column densities 
for the three isomers, in the range $(1.5-3)\times10^{12}$\,cm$^{-2}$. Our chemical model 
provides abundances of the order of the observed ones, although it overestimates the abundance 
of CH$_3$CCCN and underestimates that of HCCCH$_2$CN. The similarity of the observed abundances 
of the three isomers suggests a common origin, most probably involving reactions of the radical 
CN with the unsaturated hydrocarbons methyl acetylene and allene. Studies of reaction kinetics 
at low temperature and further observations of these molecules in different astronomical 
sources are needed to draw a clear picture of the chemistry of C$_4$H$_3$N isomers in space.}


\keywords{Astrochemistry --
		ISM: abundances --
                ISM: clouds, TMC-1 --
                ISM: molecules --
		line: identification
               }

\titlerunning{C$_4$H$_3$N isomers in TMC-1}

\maketitle
%

\section{Introduction}

Three C$_4$H$_3$N isomers have been detected in space to date. These are, in order of 
increasing energy, methylcyanoacetylene (CH$_3$C$_3$N), cyanoallene (CH$_2$CCHCN), and 
propargyl cyanide (HCCCH$_2$CN). Our knowledge of C$_4$H$_3$N isomers in the interstellar 
medium is the result of a nice multidisciplinary story with contributions from theoretical 
calculations, laboratory experiments, and astronomical observations. The presence of 
cyanoallene in cold interstellar clouds was predicted by \cite{Balucani2000,Balucani2002} 
based on crossed molecular beam experiments and \emph{ab initio} calculations which indicated 
that the reaction of CN and CH$_3$CCH would produce CH$_3$C$_3$N, already detected in TMC-1 
\citep{Broten1984}, and CH$_2$CCHCN in nearly equal amounts. Laboratory experiments indeed 
showed that the reaction CN + CH$_3$CCH is rapid at low temperatures \citep{Carty2001}. 
These results motivated an astronomical search for cyanoallene in TMC-1, which turned out 
to be successful using the GBT \citep{Lovas2006} and Effelsberg 100m \citep{Chin2006} 
telescopes.


In their combined crossed beam and \emph{ab initio} study, \cite{Balucani2000,Balucani2002} 
studied also the reaction between CN and CH$_2$CCH$_2$ (allene), a non polar metastable isomer 
of CH$_3$CCH which is thought to be also present in cold interstellar clouds. 
These authors found that the reaction should be rapid at low temperatures, something that 
was confirmed by \cite{Carty2001}, producing cyanoallene and the third C$_4$H$_3$N isomer: 
HCCCH$_2$CN. This isomer was not detected in TMC-1 by \cite{Lovas2006}, although it was later 
on found toward this same source during a cm line survey with the GBT \citep{McGuire2020}. 
The detection of propargyl cyanide in TMC-1 by these authors relied on four individual lines 
detected at a modest signal-to-noise ratio (SNR) and was supported by line stacking of 68 
transitions.

Here we present an independent and robust detection of HCCCH$_2$CN in TMC-1, with 10 lines 
detected with SNR above 10 plus 12 lines detected above 3$\sigma$, together with observations 
of the two other C$_3$H$_4$N isomers, CH$_3$C$_3$N and CH$_2$CCHCN. 
The presence of the latter is confirmed by the detection of a significant number of rotational lines.
The high sensitivity and number of lines detected allow us to derive precise abundances for 
the three isomers in a coherent and systematic way and to revisit the chemistry of C$_3$H$_4$N 
isomers in TMC-1.

\section{Observations}

The data presented here are part of a deep spectral line survey in the Q band toward TMC-1, performed
at the Yebes 40\,m radiotelescope\footnote{\texttt{http://rt40m.oan.es/rt40m$\_$en.php}} 
\citep{deVicente2016}, located at 990\,m of altitude near Guadalajara (Spain).
The observed position corresponds to the cyanopolyyne peak in TMC-1, at 
$\alpha_{J2000}=4^{\rm h} 41^{\rm  m} 41.9^{\rm s}$ and $\delta_{J2000}=+25^\circ 41' 27.0''$.
We have covered the full Q band at the 40\,m telescope, between 31.1\,GHz and 50.4\,GHz, 
using the recently installed NANOCOSMOS HEMT Q band receiver \citep{TerceroF2020} and the
fast Fourier transform spectrometers (FFTS) with 8$\times$2.5\,GHz bands per lineal polarization,
which allow a simultaneous scan of a band width of 18\,GHz at a spectral resolution of 38\,kHz ($\sim$0.27\,km\,s$^{-1}$).
We observed two setups at different central frequencies in order to fully cover the lower 
and upper frequencies allowed by the Q band receiver, and to check for spurious signals and 
other technical artifacts.

The observations were performed in several sessions, between November 2019 and February 2020,
using the frequency switching technique with a frequency throw of 10\,MHz.
The intensity scale in the spectra obtained is T$_{\rm A}^*$, antenna temperature corrected for 
atmospheric absorption and spillover losses, which
was calibrated using two absorbers at different temperatures and the 
atmospheric transmission model ATM \citep{Cernicharo1985,Pardo2001}.
Pointing and focus were checked every hour through pseudo-continuum observations (see e.g. 
\citealt{deVicente2016, Tercero2020}) of the SiO $J=1-0$, $v=1$ maser emission towards the 
O-rich evolved star IK Tau, which is close to the target source. The pointing errors were always found within 2-3$''$.
System temperatures were in the range 50-250\,K 
depending on the frequency, the particular weather conditions of each observing session
(from 5\,mm to 10\,mm of precipitable water vapor), and the elevation of the source (from 15$^{\circ}$ to 80$^{\circ}$).
The final rms obtained is in the range 0.5-1\,mK, rising up to 3\,mK at the highest frequencies.
The main beam efficiency of the Yebes 40\,m telescope ranges from 0.6 at 32\,GHz to 0.43 at 49\,GHz, and the half power beam width (HPBW) ranges from 55$''$ at 32\,GHz to 37$''$ at 49\,GHz. All the data were reduced and analyzed using the 
\texttt{GILDAS}\footnote{\texttt{http://www.iram.fr/IRAMFR/GILDAS/}} software.

\section{Results}

\begin{figure*}
\centering
\includegraphics[width=0.95\textwidth]{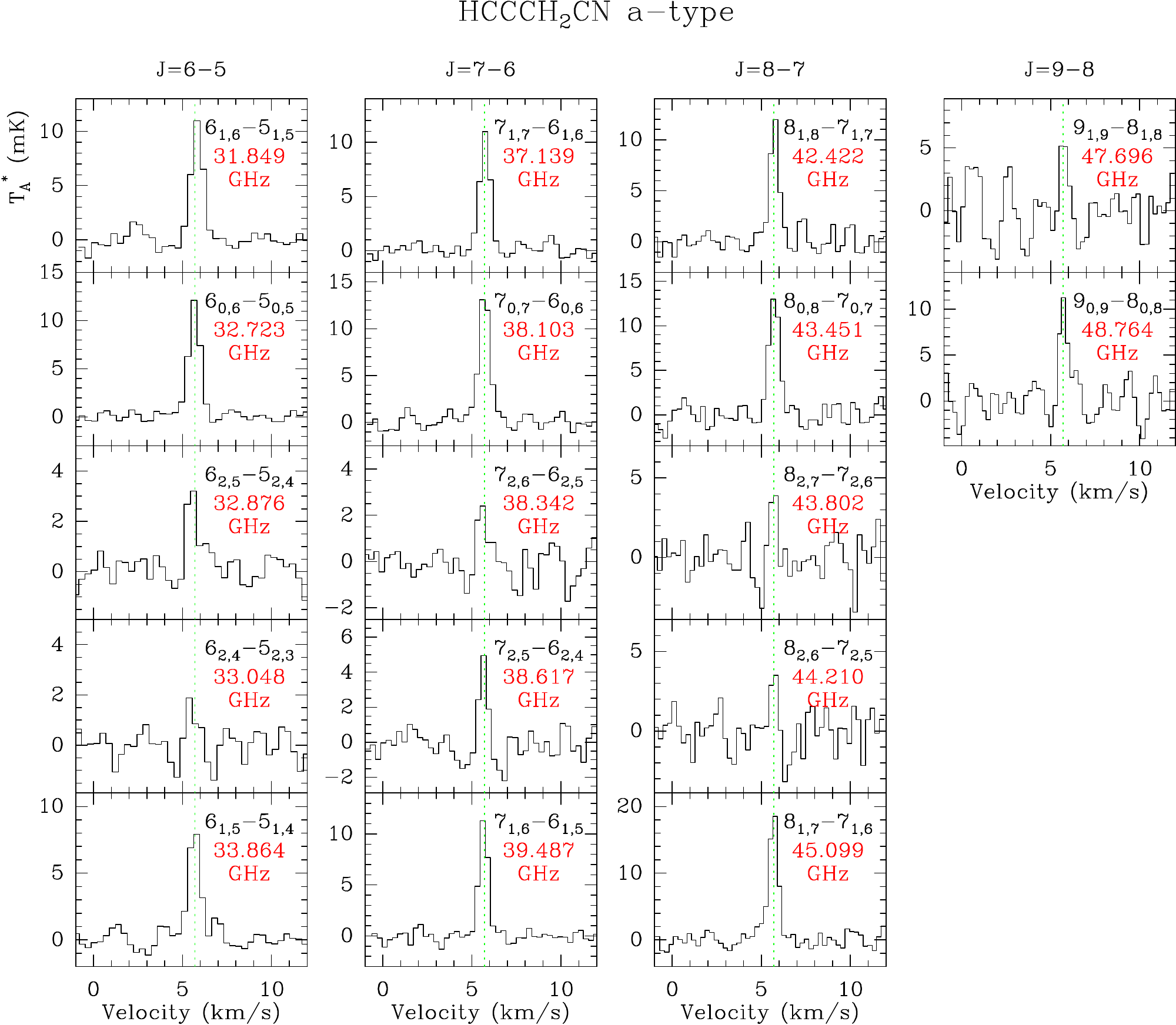}
\caption{Observed lines of HCCCH$_2$CN ($a$-type) toward TMC-1 (CP). The vertical dashed green line marks a radial velocity 
of 5.7\,km\,s$^{-1}$.}
\label{fig:lines-HCCCH2CN}
\end{figure*}

\begin{figure}
\centering
\includegraphics[width=0.70\columnwidth]{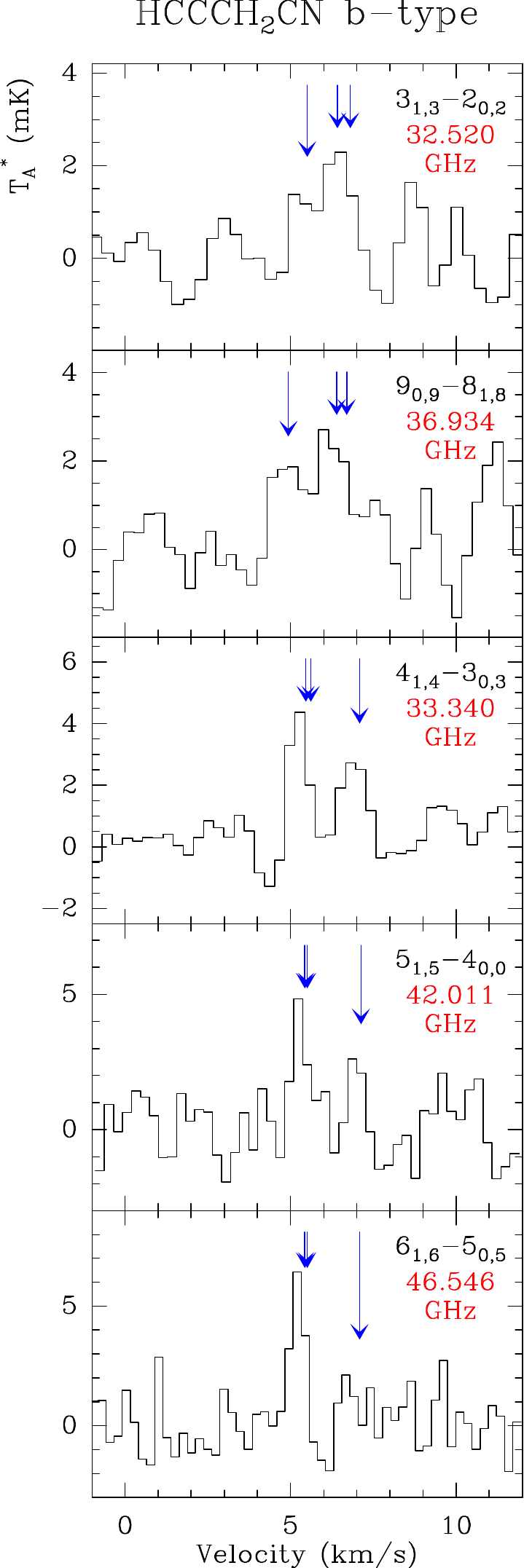}
\caption{Observed lines of HCCCH$_2$CN ($b$-type) toward TMC-1 (CP).
Blue arrows show the position of the strongest three hyperfine components. 
Velocity axis refers to the frequency result of collapsing the hyperfine structure.}
\label{fig:lines-HCCCH2CN-btype}
\end{figure}

The high sensitivity of this line survey allowed the detection
of HCCCH$_2$CN towards TMC-1 through 17 $a$-type lines up to quantum numbers $J=9-8$
and $K_{\rm a}=0,1,2$ ($E_{\rm u}\leq13$\,K), with 10 of them showing a SNR $>$10. 
In addition, we detected 10 $b$-type lines harbouring hyperfine structure. 
These lines are shown in Fig.\,\ref{fig:lines-HCCCH2CN} and Fig.\,\ref{fig:lines-HCCCH2CN-btype}
and are listed in Table\,\ref{table:fits}.
Line identification was performed using the \texttt{MADEX} catalogue (\citealt{MADEX}, see Table\,\ref{table:fits}), 
which also includes predictions for the hyperfine structure.
This detection confirms the presence of this species in space, recently claimed for the 
first time in TMC-1 by \citet{McGuire2020} using the Green Bank Telescope (GBT). These 
authors presented a 5$\sigma$ signal (18$\sigma$ in the response impulse function) obtained by 
an intensity and noise-weighted average (``stack'') of the data at the expected frequencies 
of the HCCCH$_2$CN lines that could be present within the noise level. 
It is worth noting that our 40\,m survey of TMC-1 in 
the Q band is complementary to that performed with the GBT between 8\,GHz and 30\,GHz.
Although most of the individual lines of HCCCH$_2$CN are below the detection limit
of the GBT data, four of them are detected at 1-3$\sigma$ levels. 
Thanks to the high spectral resolution of these data (1.4\,kHz) they
distinguished three cloud components in the line profiles (see \citealt{Fosse2001}
for a detailed analysis of the velocity structure of this source).

In this work, a single Gaussian function was fitted to the HCCCH$_2$CN line profiles
to obtain the observed line parameters (see Table\,\ref{table:fits}).
We derived a $V_{\rm LSR}=(5.70\pm0.09)$\,km\,s$^{-1}$ and
a line width ($\Delta$$v$, full width at half maximum) of $(0.66\pm0.18)$\,km\,s$^{-1}$.
The former is slightly different from the value $(5.83\pm0.01)$\,km\,s$^{-1}$,
obtained by \citet{Cernicharo2020a} from 
Gaussian fits to the 50 lines of HC$_5$N and its $^{13}$C and $^{15}$N isotopologues 
detected in our line survey. 
Note we have a larger uncertainty due to the lower number of transitions and the weakness 
of some of the lines as compared to HC$_5$N, in particular the $b$-type transitions.

We also detected the other two C$_4$H$_3$N isomers, CH$_2$CCHCN and CH$_3$CCCN, 
using frequencies from the \texttt{MADEX} catalogue (\citealt{MADEX}, see Table\,\ref{table:fits}).
The 16 lines of CH$_2$CCHCN detected in our line survey are shown in Fig.\,\ref{fig:lines-CH2CCHCN} and are listed in Table\,\ref{table:fits}. All of them are detected above a 10$\sigma$ level. 
This species was previously identified in TMC-1 through four lines between 20\,GHz and 26\,GHz \citep{Lovas2006}. Here we report the first 
detection of lines of CH$_2$CCHCN above 30\,GHz in TMC-1.
\citet{Kaifu2004} did not detect lines above the noise limit at the CH$_2$CCHCN frequencies in their line survey between 8.8\,GHz and 50\,GHz carried out with the Nobeyama 45\,m telescope.
As we mentioned in previous works \citep{Cernicharo2020a,Cernicharo2020b,Cernicharo2020c,Marcelino2020}, 
the sensitivity of our observations is a factor 5-10 better than
that of \citet{Kaifu2004} at the same frequencies.
The derived $V_{\rm LSR}$ for the CH$_2$CCHCN lines, by fitting a single Gaussian, is 
$(5.66\pm0.03)$\,km\,s$^{-1}$, which is similar, within errors, to the one obtained for HCCCH$_2$CN. 
The isomer CH$_3$CCCN, a well known species in TMC-1 \citep{Broten1984,Kaifu2004}, has been also identified in our line survey through 10 strong lines ($J_{\rm u}$ from 8 to 12 and $K=0,1$) plus five $K=2$ lines ($E_{\rm u}>29$\,K) tentatively detected (see Fig.\,\ref{fig:lines-CH3CCCN} and Table\,\ref{table:fits}). These lines show a $V_{\rm LSR}$ of 
$(5.80\pm0.02)$\,km\,s$^{-1}$ which matches that observed for HC$_5$N.

We can estimate rotational temperatures ($T_{\rm rot}$)
and molecular column densities ($N$) for the detected species by constructing
rotational diagrams (see e.g. \citealt{Goldsmith1999}).
This analysis assumes the Rayleigh-Jeans approximation, optically thin lines,
and LTE conditions. 
The equation that derives the total column density under these conditions
can be re-arranged as

\begin{equation}
{\rm \ln} \left(\frac{8 \pi k_{\rm B} \nu^2 \int{T_{\rm MB} dv}}{h c^3 A_{\rm ul} g_{\rm u} b}\right) = {\rm \ln} \left(\frac{N}{Q_{\rm rot}} \frac{T_{\rm rot}-T_{\rm bg}}{T_{\rm rot}}\right) - \frac{E_{\rm u}}{k_{\rm B} T_{\rm rot}}
\label{eq_RD}
,\end{equation}
where $g_u$ is the statistical weight in the upper level,
$A_{\rm ul}$ is the Einstein $A$-coefficient for
spontaneous emission, $Q_{\rm rot}$ is the rotational partition
function which depends on $T_{\rm rot}$,
$E_{\rm u}$ is the upper level energy, $\nu$ is the frequency
of the transition, $b$ is the dilution factor, and $T_{\rm bg}$ is the cosmic microwave background radiation temperature.
We assume a source diameter of 80$''$ (see \citealt{Fosse2001}). The first term of Eq.\,(\ref{eq_RD}), which depends only on spectroscopic and
observational line parameters, is plotted as a function of $E_{\rm u}$/$k_{\rm B}$
for the different lines detected. Thus, $T_{\rm rot}$ and $N$ can be derived
by performing a linear least squares fit to the points (see Fig.\,\ref{fig_DR}).

Results for $T_{\rm rot}$ and $N$ using the population diagram procedure are shown in 
Table\,\ref{table_cd} and Fig.\,\ref{fig_DR}.
The uncertainties were calculated using the statistical
errors given by the linear least squares fit for the slope and the
intercept. The individual errors of the data points are
those derived by taking into account the uncertainty obtained in the determination 
of the observed line parameters (see Table\,\ref{table:fits}).
For HCCCH$_2$CN ($a$-type transitions) and CH$_2$CCHCN, different hyperfine structure components of the same 
$(J_{K_{\rm a},K_{\rm c}})_{\rm u}-(J_{K_{\rm a},K_{\rm c}})_{\rm l}$
transition are blended in a single line.
Thus, to correctly determine $T_{\rm rot}$ and $N$,  
the Einstein $A$-coefficient for
spontaneous emission and the statistical weight were assumed as the weighted average values
of the sum of the hyperfine components, and the rotational partition
function was calculated using this value for the statistical weight of each 
$(J_{K_{\rm a},K_{\rm c}})_{\rm u}-(J_{K_{\rm a},K_{\rm c}})_{\rm l}$ transition.
For CH$_3$CCCN we built independent rotational diagrams for each symmetry state $A$ and $E$.

We obtained rotational temperatures 
between $4-8$\,K for the three isomers (see Table\,\ref{table_cd}), 
indicating they are subthermally excited, like most of the species in this region 
(see e.g. \citealt{Cernicharo2020a, Cernicharo2020c, Marcelino2020}). 
For the column density, we derived very similar values of the three isomers,
in the range $(1.5-3)\times10^{12}$\,cm$^{-2}$.

\begin{table}
\begin{center}
\caption{Derived rotational temperatures ($T_{\rm rot}$) and column densities ($N$)
for the C$_4$H$_3$N isomers towards TMC-1 (CP).}
\label{table_cd}
\begin{tabular}{ccc}   
\hline
Species & $T_{\rm rot}$ (K) & $N$ (cm$^{-2}$) \\
\hline
\hline
HCCCH$_2$CN & $4\pm1$ & $(2.8\pm0.7)\times10^{12}$\\
CH$_2$CCHCN & $5.5\pm0.3$ & $(2.7\pm0.2)\times10^{12}$\\
A-CH$_3$CCCN & $6.7\pm0.2$ & $(9.7\pm0.3)\times10^{11}$\\
E-CH$_3$CCCN & $8.2\pm0.6$ & $(7.7\pm0.5)\times10^{11}$\\
\hline
\end{tabular}
\end{center}
\end{table}

\section{Discussion}

The chemistry of C$_4$H$_3$N isomers in cold molecular clouds has been discussed by 
\cite{Balucani2000} and more specifically by \cite{Balucani2002}, 
based on crossed molecular beam experiments and \emph{ab initio} calculations.
In these studies it was pointed out that reactions of the CN radical with methyl acetylene 
and allene are barrierless and exothermic when producing CH$_3$C$_3$N and CH$_2$CCHCN, in 
the methyl acetylene reaction, and CH$_2$CCHCN and HCCCH$_2$CN, in the reaction involving 
allene. Indeed, the reactions of CN with CH$_3$CCH and CH$_2$CCH$_2$ were measured to be 
rapid at low temperatures \citep{Carty2001}. This chemical scheme 
was implemented in a chemical model by \cite{Quan2007} to explain the abundance of 
cyanoallene in TMC-1. Later on, \cite{Abeysekera2015} measured the product branching 
ratios of the reaction between CN and methyl acetylene at low temperature 
using a chirped-pulse uniform flow and found that HC$_3$N is the major product, while 
CH$_3$C$_3$N accounts for 22~\% of the products and CH$_2$CCHCN is not formed. 
These results are in contrast with those obtained from crossed molecular beam experiments 
\citep{Huang1999,Balucani2000,Balucani2002}, where CH$_2$CCHCN is observed as product of 
the CN + CH$_3$CCH reaction. Therefore, the most stable isomer CH$_3$C$_3$N can be formed 
in the reaction of CN and methyl acetylene, the second most stable isomer CH$_2$CCHCN can 
be formed when CN reacts with CH$_2$CCH$_2$ and perhaps also with CH$_3$CCH, depending on 
whether one gives credit to the chirped-pulse uniform flow experiment or to the crossed 
molecular beam ones, and the least stable isomer HCCCH$_2$CN can only be formed in the 
reaction between CN and allene. 
These neutral-neutral reactions involving CN are therefore likely routes to the three 
C$_4$H$_3$N isomers in cold interstellar clouds like TMC-1, where abundant CN, CH$_3$CCH, 
and probably CH$_2$CCH$_2$ (non polar and thus it cannot be detected at radio wavelengths) 
are present. 
Moreover, the presence of HCCCH$_2$CN (and perhaps also CH$_2$CCHCN) can be used as 
proxy of the non polar C$_3$H$_4$ isomer allene since this isomer is only formed from 
CH$_2$CCH$_2$ in the aforementioned reactions of CN.

In the light of the recent discovery of HCCCH$_2$CN in TMC-1 and the observational study 
of the three C$_4$H$_3$N isomers presented here, we have carried out chemical model 
calculations to review the chemistry of these species in cold clouds and evaluate whether 
the mechanism proposed by \cite{Balucani2002} is in agreement with observations. We adopt 
typical parameters of cold dark clouds, i.e., a gas kinetic temperature of 10 K, a volume 
density of H nuclei of $2\times10^4$ cm$^{-3}$, a visual extinction of 30 mag, a cosmic-ray 
ionization rate of H$_2$ of $1.3\times10^{-17}$ s$^{-1}$, and the so-called "low-metal" 
elemental abundances \citep{Agundez2013}. We use the chemical network {\small RATE12} from 
the UMIST database \citep{McElroy2013}, updated to include the C$_4$H$_3$N isomers 
CH$_2$CCHCN and HCCCH$_2$CN. The reactions
\begin{subequations} \label{reac:cn+ch3cch}
\begin{align}
\rm CN + CH_3CCH & \rightarrow \rm HCN + CH_2CCH, \label{reac:cn+ch3cch_a} \\
                                 & \rightarrow \rm HC_3N + CH_3, \label{reac:cn+ch3cch_b} \\
                                 & \rightarrow \rm CH_3C_3N + H, \label{reac:cn+ch3cch_c} \\
                                 & \rightarrow \rm CH_2CCHCN + H, \label{reac:cn+ch3cch_d}
\end{align}
\end{subequations}
\begin{subequations} \label{reac:cn+ch2cch2}
\begin{align}
\rm CN + CH_2CCH_2 & \rightarrow \rm CH_2CCHCN + H, \label{reac:cn+ch2cch2_a} \\
                                      & \rightarrow \rm HCCCH_2CN + H, \label{reac:cn+ch2cch2_b}
\end{align}
\end{subequations}
are included with the rate constants measured by \cite{Carty2001}. 
For the branching ratios of reaction (\ref{reac:cn+ch3cch}) we use either the values 
measured in the chirped-pulse uniform flow experiment by \cite{Abeysekera2015}, 12~\%, 66~\%, 
22~\%, and 0~\% for channels (a), (b), (c), and (d), respectively, or the values suggested 
by crossed molecular beam experiments and quantum chemical calculations \cite{Balucani2000}, 
50~\% for channels (c) and (d). For reaction (\ref{reac:cn+ch2cch2}) we adopt branching 
ratios of 90~\% and 10~\% for channels (a) and (b), respectively, based on quantum chemical 
calculations by \cite{Balucani2002}. The destruction processes of CH$_2$CCHCN and HCCCH$_2$CN 
are assumed to be the same as those of CH$_3$C$_3$N, which are basically reactions with 
abundant cations.

The calculated abundances of the three C$_4$H$_3$N isomers are shown as a function of time 
in Fig.~\ref{fig:abun}. It is seen that the three isomers reach their maximum abundance at 
early times, in the range $(1-4)\times10^5$ yr, with CH$_3$C$_3$N being the most abundant 
and HCCCH$_2$CN being the least abundant. According to the chemical model, the formation 
of CH$_3$C$_3$N occurs through two routes. The first and major involves the dissociative 
recombination of the precursor ion CH$_3$C$_3$NH$^+$ with electrons and is the responsible 
of the larger calculated abundance of CH$_3$C$_3$N compared to the two other isomers. A 
second and minor route is provided by reaction~(\ref{reac:cn+ch3cch_c}). 
Cyanoallene is formed through reaction (\ref{reac:cn+ch2cch2}), with reaction~(\ref{reac:cn+ch3cch_c}) 
contributing to the same level if channel~(\ref{reac:cn+ch3cch_d}) is assumed to be open. 
Propargyl cyanide is exclusively formed through reaction (\ref{reac:cn+ch2cch2}), with a 
lower abundance because it is formed with a branching ratio of just 10~\%. The impact of 
using the branching ratios for reaction~(\ref{reac:cn+ch3cch}) of \cite{Balucani2000} or 
those of \cite{Abeysekera2015} is modest, with the main effect being a change of less than 
a factor of two in the abundance of CH$_2$CCHCN (see Fig.~\ref{fig:abun}).

The fact that the observed abundances of the three isomers are remarkably similar provides clues on the underlying chemical processes at work. For example, the route to CH$_3$C$_3$N from the precursor ion CH$_3$C$_3$NH$^+$ is probably overestimated in the chemical model, as indicated by the too large abundance calculated for this species. It has become clear in recent years that dissociative recombination of polyatomic ions usually results in a much larger fragmentation than previously believed \citep{Larsson2012}, meaning that it would not be strange than CH$_3$C$_3$N is a minor product in the dissociative recombination of CH$_3$C$_3$NH$^+$. The low branching ratio adopted for HCCCH$_2$CN formation in reaction (\ref{reac:cn+ch2cch2}) based on calculations by \cite{Balucani2002} seems also to be in conflict with the observational finding of similar abundances for CH$_2$CCHCN and HCCCH$_2$CN. It would be very interesting to measure the product branching ratios for the reaction of CN with allene, as was done for CN + CH$_3$CCH \citep{Abeysekera2015}, to shed light on the formation routes of these two metastable C$_4$H$_3$N isomers. This will also allow to put tight constraints on the abundance of allene in cold dense clouds.

In summary, the similar abundances observed for the three C$_4$H$_3$N isomers favors a common origin through reactions (\ref{reac:cn+ch3cch}) and (\ref{reac:cn+ch2cch2}) with similar branching ratios in the latter reaction. If this scenario is correct, we can conclude that allene is as abundant as methyl acetylene in TMC-1. This is in fact predicted by the chemical model, where CH$_3$CCH and CH$_2$CCH$_2$ are mostly formed during the dissociative recombination of the C$_3$H$_7^+$ ion \citep{Larsson2005}, with similar branching ratios assumed for the two C$_3$H$_4$ isomers.

In addition to the three C$_4$H$_3$N isomers and the well known species HC$_3$N and CH$_2$CHCN, \citet{Balucani2000}
predicted the presence of $c$-C$_6$H$_5$CN and the C$_5$H$_5$N isomer CH$_2$CC(CN)CH$_3$ in cold interstellar clouds. 
It is worth noting
that all these species but CH$_2$CC(CN)CH$_3$ have been identified in TMC-1 (see \citealt{McGuire2018}
for the detection of cyanobenzene) and are also present in our survey. Another $-$CN species,
cyanocyclopentadiene ($c$-C$_5$H$_5$CN), has been recently detected in this source \citep{McCarthy2020}.
A complete study of the molecular rings $c$-C$_6$H$_5$CN and $c$-C$_5$H$_5$CN in our data
will be published elsewhere. We searched in our data for the two C$_5$H$_5$N isomers CH$_3$CH$_2$CCCN and CH$_3$CHCCHCN by performing a line stacking analysis (see, e.g., \citealt{Cuadrado2016,Loomis2020}).
We added spectra at the expected frequency of several lines from these species that could be present within the noise level. More concreteley, we considered $a$-type transitions sharing similar upper level energies, up to 15\,K, and Einstein coefficients.
All spectra, in local standard of rest (LSR) velocity scale, are resampled to the same velocity
channel resolution before stacking. Figure\,\ref{fig_stack} shows the spectra obtained following
this method. Whereas there is no evidence for the presence of CH$_3$CH$_2$CCCN in our data,
the stacked spectrum of CH$_3$CHCCHCN shows a 2$\sigma$ signal at the systemic velocity of the source.
An observational effort at lowest frequencies has to be undertaken to confirm the
presence of CH$_3$CHCCHCN in space.

\section{Conclusions}

Using a very sensitive line survey of TMC-1 in the Q band we have detected multiple transitions of the 
three C$_4$H$_3$N isomers CH$_3$C$_3$N, CH$_2$CCHCN, and HCCCH$_2$CN. The presence of
the latter in TMC-1 is supported by 27 observed individual lines.
We have constructed rotational diagrams for the three species and obtained similar 
rotational temperatures and column densities for the three isomers, 
in the range of $4-8$\,K and $(1.5-3)\times10^{12}$\,cm$^{-2}$, respectively.
The observed abundances of the three isomers in TMC-1 suggest a similar chemical origin based 
on reactions of the radical CN with the isomers CH$_3$CCH and CH$_2$CCH$_2$. There are still 
uncertainties in the network of reactions related to these species since our chemical model 
overestimates the abundance of CH$_3$C$_3$N and underestimates the production of HCCCH$_2$CN. 
Further studies of these isomers in other sources could help in clarifying their chemical 
formation pathways.

\begin{acknowledgements}
We acknowledge funding support from the European Research Council (ERC Grant 610256: NANOCOSMOS). We also thank the Spanish MICIU for funding support under grants AYA2016-75066-C2-1-P, PID2019-106110GB-I00, and PID2019-107115GB-C21, and PID2019-106235GB-I00. M.A. thanks MICIU for grant RyC-2014-16277.
\end{acknowledgements}

%
%

\bibliographystyle{aa}
\bibliography{references_c4h3n}

\begin{thebibliography}{39}
\expandafter\ifx\csname natexlab\endcsname\relax\def\natexlab#1{#1}\fi

\bibitem[{Abeysekera {et~al.}(2015)Abeysekera, Joalland, Ariyasingha, Zack,
  Sims, Field, \& Suits}]{Abeysekera2015}
Abeysekera, C., Joalland, B., Ariyasingha, N., {et~al.} 2015, The Journal of
  Physical Chemistry Letters, 6, 1599

\bibitem[{{Ag{\'u}ndez} \& {Wakelam}(2013)}]{Agundez2013}
{Ag{\'u}ndez}, M. \& {Wakelam}, V. 2013, Chemical Reviews, 113, 8710

\bibitem[{{Balucani} {et~al.}(2000){Balucani}, {Asvany}, {Huang}, {Lee},
  {Kaiser}, {Osamura}, \& {Bettinger}}]{Balucani2000}
{Balucani}, N., {Asvany}, O., {Huang}, L.~C.~L., {et~al.} 2000, \apj, 545, 892

\bibitem[{{Balucani} {et~al.}(2002){Balucani}, {Asvany}, {Kaiser}, \&
  {Osamura}}]{Balucani2002}
{Balucani}, N., {Asvany}, O., {Kaiser}, R.~I., \& {Osamura}, Y. 2002, Journal
  of Physical Chemistry A, 106, 4301

\bibitem[{{Bester} {et~al.}(1983){Bester}, {Tanimoto}, {Vowinkel},
  {Winnewisser}, \& {Yamada}}]{Bester1983}
{Bester}, M., {Tanimoto}, M., {Vowinkel}, B., {Winnewisser}, G., \& {Yamada},
  K. 1983, Zeitschrift Naturforschung Teil A, 38, 64

\bibitem[{{Bester} {et~al.}(1984){Bester}, {Yamada}, {Winnewisser}, {Joentgen},
  {Altenbach}, \& {Vogel}}]{Bester1984}
{Bester}, M., {Yamada}, K., {Winnewisser}, G., {et~al.} 1984, \aap, 137, L20

\bibitem[{{Bouchy} {et~al.}(1973){Bouchy}, {Demaison}, {Roussy}, \&
  {Barriol}}]{Bouchy1973}
{Bouchy}, A., {Demaison}, J., {Roussy}, G., \& {Barriol}, J. 1973, Journal of
  Molecular Structure, 18, 211

\bibitem[{{Broten} {et~al.}(1984){Broten}, {MacLeod}, {Avery}, {Irvine},
  {Hoglund}, {Friberg}, \& {Hjalmarson}}]{Broten1984}
{Broten}, N.~W., {MacLeod}, J.~M., {Avery}, L.~W., {et~al.} 1984, \apjl, 276,
  L25

\bibitem[{{Carty} {et~al.}(2001){Carty}, {Le Page}, {Sims}, \&
  {Smith}}]{Carty2001}
{Carty}, D., {Le Page}, V., {Sims}, I.~R., \& {Smith}, I. W.~M. 2001, Chemical
  Physics Letters, 344, 310

\bibitem[{{Cernicharo}(1985)}]{Cernicharo1985}
{Cernicharo}, J. 1985, Internal IRAM Report (Granada: IRAM)

\bibitem[{{Cernicharo}(2012)}]{MADEX}
{Cernicharo}, J. 2012, in EAS Publications Series, Vol.~58, EAS Publications
  Series, ed. C.~{Stehl{\'e}}, C.~{Joblin}, \& L.~{d'Hendecourt}, 251--261

\bibitem[{{Cernicharo} \& {Guelin}(1987)}]{Cernicharo1987}
{Cernicharo}, J. \& {Guelin}, M. 1987, \aap, 176, 299

\bibitem[{{Cernicharo} {et~al.}(2020{\natexlab{a}}){Cernicharo}, {Marcelino},
  {Ag{\'u}ndez}, {Berm{\'u}dez}, {Cabezas}, {Tercero}, \&
  {Pardo}}]{Cernicharo2020a}
{Cernicharo}, J., {Marcelino}, N., {Ag{\'u}ndez}, M., {et~al.}
  2020{\natexlab{a}}, \aap, 642, L8

\bibitem[{{Cernicharo} {et~al.}(2020{\natexlab{b}}){Cernicharo}, {Marcelino},
  {Ag{\'u}ndez}, {Endo}, {Cabezas}, {Berm{\'u}dez}, {Tercero}, \& {de
  Vicente}}]{Cernicharo2020b}
{Cernicharo}, J., {Marcelino}, N., {Ag{\'u}ndez}, M., {et~al.}
  2020{\natexlab{b}}, \aap, 642, L17

\bibitem[{{Cernicharo} {et~al.}(2020{\natexlab{c}}){Cernicharo}, {Marcelino},
  {Pardo}, {Ag{\'u}ndez}, {Tercero}, {de Vicente}, {Cabezas}, \&
  {Berm{\'u}dez}}]{Cernicharo2020c}
{Cernicharo}, J., {Marcelino}, N., {Pardo}, J.~R., {et~al.} 2020{\natexlab{c}},
  \aap, 641, L9

\bibitem[{{Chin} {et~al.}(2006){Chin}, {Kaiser}, {Lemme}, \&
  {Henkel}}]{Chin2006}
{Chin}, Y.-N., {Kaiser}, R.~I., {Lemme}, C., \& {Henkel}, C. 2006, in American
  Institute of Physics Conference Series, Vol. 855, Astrochemistry - From
  Laboratory Studies to Astronomical Observations, ed. R.~I. {Kaiser},
  P.~{Bernath}, Y.~{Osamura}, S.~{Petrie}, \& A.~M. {Mebel}, 149--153

\bibitem[{{Cuadrado} {et~al.}(2016){Cuadrado}, {Goicoechea}, {Roncero},
  {Aguado}, {Tercero}, \& {Cernicharo}}]{Cuadrado2016}
{Cuadrado}, S., {Goicoechea}, J.~R., {Roncero}, O., {et~al.} 2016, \aap, 596,
  L1

\bibitem[{{de Vicente} {et~al.}(2016){de Vicente}, {Bujarrabal},
  {D{\'\i}az-Pulido}, {Albo}, {Alcolea}, {Barcia}, {Barbas}, {Bola{\~n}o},
  {Colomer}, {Diez}, {Gallego}, {G{\'o}mez-Gonz{\'a}lez},
  {L{\'o}pez-Fern{\'a}ndez}, {L{\'o}pez-Fern{\'a}ndez}, {L{\'o}pez-P{\'e}rez},
  {Malo}, {Moreno}, {Patino}, {Serna}, {Tercero}, \& {Vaquero}}]{deVicente2016}
{de Vicente}, P., {Bujarrabal}, V., {D{\'\i}az-Pulido}, A., {et~al.} 2016,
  \aap, 589, A74

\bibitem[{{Demaison} {et~al.}(1985){Demaison}, {Pohl}, \&
  {Rudolph}}]{Demaison1985}
{Demaison}, J., {Pohl}, I., \& {Rudolph}, H.~D. 1985, Journal of Molecular
  Spectroscopy, 114, 210

\bibitem[{{Foss{\'e}} {et~al.}(2001){Foss{\'e}}, {Cernicharo}, {Gerin}, \&
  {Cox}}]{Fosse2001}
{Foss{\'e}}, D., {Cernicharo}, J., {Gerin}, M., \& {Cox}, P. 2001, \apj, 552,
  168

\bibitem[{{Goldsmith} \& {Langer}(1999)}]{Goldsmith1999}
{Goldsmith}, P.~F. \& {Langer}, W.~D. 1999, \apj, 517, 209

\bibitem[{Huang {et~al.}(1999)Huang, Balucani, Lee, Kaiser, \&
  Osamura}]{Huang1999}
Huang, L. C.~L., Balucani, N., Lee, Y.~T., Kaiser, R.~I., \& Osamura, Y. 1999,
  The Journal of Chemical Physics, 111, 2857

\bibitem[{{Kaifu} {et~al.}(2004){Kaifu}, {Ohishi}, {Kawaguchi}, {Saito},
  {Yamamoto}, {Miyaji}, {Miyazawa}, {Ishikawa}, {Noumaru}, {Harasawa}, {Okuda},
  \& {Suzuki}}]{Kaifu2004}
{Kaifu}, N., {Ohishi}, M., {Kawaguchi}, K., {et~al.} 2004, \pasj, 56, 69

\bibitem[{{Larsson} {et~al.}(2005){Larsson}, {Ehlerding}, {Geppert},
  {Hellberg}, {Kalhori}, {Thomas}, {Djuric}, {{\"O}sterdahl}, {Angelova},
  {Semaniak}, {Novotny}, {Arnold}, \& {Viggiano}}]{Larsson2005}
{Larsson}, M., {Ehlerding}, A., {Geppert}, W.~D., {et~al.} 2005, \jcp, 122,
  156101

\bibitem[{{Larsson} {et~al.}(2012){Larsson}, {Geppert}, \&
  {Nyman}}]{Larsson2012}
{Larsson}, M., {Geppert}, W.~D., \& {Nyman}, G. 2012, Reports on Progress in
  Physics, 75, 066901

\bibitem[{{Loomis} {et~al.}(2020){Loomis}, {Burkhardt}, {Shingledecker},
  {Charnley}, {Cordiner}, {Herbst}, {Kalenskii}, {Lee}, {Willis}, {Xue},
  {Remijan}, {McCarthy}, \& {McGuire}}]{Loomis2020}
{Loomis}, R.~A., {Burkhardt}, A.~M., {Shingledecker}, C.~N., {et~al.} 2020,
  arXiv e-prints, arXiv:2009.11900

\bibitem[{{Lovas} {et~al.}(2006){Lovas}, {Remijan}, {Hollis}, {Jewell}, \&
  {Snyder}}]{Lovas2006}
{Lovas}, F.~J., {Remijan}, A.~J., {Hollis}, J.~M., {Jewell}, P.~R., \&
  {Snyder}, L.~E. 2006, \apjl, 637, L37

\bibitem[{{Marcelino} {et~al.}(2020){Marcelino}, {Ag{\'u}ndez}, {Tercero},
  {Cabezas}, {Berm{\'u}dez}, {Gallego}, {deVicente}, \&
  {Cernicharo}}]{Marcelino2020}
{Marcelino}, N., {Ag{\'u}ndez}, M., {Tercero}, B., {et~al.} 2020, \aap, 643, L6

\bibitem[{{McCarthy} {et~al.}(2020){McCarthy}, {Lee}, {Loomis}, {Burkhardt},
  {Shingledecker}, {Charnley}, {Cordiner}, {Herbst}, {Kalenskii}, {Willis},
  {Xue}, {Remijan}, \& {McGuire}}]{McCarthy2020}
{McCarthy}, M.~C., {Lee}, K. L.~K., {Loomis}, R.~A., {et~al.} 2020, Nature
  Astronomy [\eprint[arXiv]{2009.13546}]

\bibitem[{{McElroy} {et~al.}(2013){McElroy}, {Walsh}, {Markwick}, {Cordiner},
  {Smith}, \& {Millar}}]{McElroy2013}
{McElroy}, D., {Walsh}, C., {Markwick}, A.~J., {et~al.} 2013, \aap, 550, A36

\bibitem[{{McGuire} {et~al.}(2018){McGuire}, {Burkhardt}, {Kalenskii},
  {Shingledecker}, {Remijan}, {Herbst}, \& {McCarthy}}]{McGuire2018}
{McGuire}, B.~A., {Burkhardt}, A.~M., {Kalenskii}, S., {et~al.} 2018, Science,
  359, 202

\bibitem[{{McGuire} {et~al.}(2020){McGuire}, {Burkhardt}, {Loomis},
  {Shingledecker}, {Kelvin Lee}, {Charnley}, {Cordiner}, {Herbst}, {Kalenskii},
  {Momjian}, {Willis}, {Xue}, {Remijan}, \& {McCarthy}}]{McGuire2020}
{McGuire}, B.~A., {Burkhardt}, A.~M., {Loomis}, R.~A., {et~al.} 2020, \apjl,
  900, L10

\bibitem[{{McNaughton} {et~al.}(1988){McNaughton}, {Romeril}, {Lappert}, \&
  {Kroto}}]{McNaughton1988}
{McNaughton}, D., {Romeril}, N.~G., {Lappert}, M.~F., \& {Kroto}, H.~W. 1988,
  Journal of Molecular Spectroscopy, 132, 407

\bibitem[{{Mo{\"\i}ses} {et~al.}(1982){Mo{\"\i}ses}, {Boucher}, {Burie},
  {Demaison}, \& {Dubrulle}}]{Moises1982}
{Mo{\"\i}ses}, A., {Boucher}, D., {Burie}, J., {Demaison}, J., \& {Dubrulle},
  A. 1982, Journal of Molecular Spectroscopy, 92, 497

\bibitem[{{Pardo} {et~al.}(2001){Pardo}, {Cernicharo}, \&
  {Serabyn}}]{Pardo2001}
{Pardo}, J.~R., {Cernicharo}, J., \& {Serabyn}, E. 2001, IEEE Transactions on
  Antennas and Propagation, 49, 1683

\bibitem[{{Quan} \& {Herbst}(2007)}]{Quan2007}
{Quan}, D. \& {Herbst}, E. 2007, \aap, 474, 521

\bibitem[{{Schwahn} {et~al.}(1986){Schwahn}, {Schieder}, {Bester}, \&
  {Winnewisser}}]{Schwahn1986}
{Schwahn}, G., {Schieder}, R., {Bester}, M., \& {Winnewisser}, G. 1986, Journal
  of Molecular Spectroscopy, 116, 263

\bibitem[{{Tercero} {et~al.}(2020{\natexlab{a}}){Tercero}, {Cernicharo},
  {Cuadrado}, {de Vicente}, \& {Gu{\'e}lin}}]{Tercero2020}
{Tercero}, B., {Cernicharo}, J., {Cuadrado}, S., {de Vicente}, P., \&
  {Gu{\'e}lin}, M. 2020{\natexlab{a}}, \aap, 636, L7

\bibitem[{{Tercero} {et~al.}(2020{\natexlab{b}}){Tercero},
  {L{\'o}pez-P{\'e}rez}, {Gallego}, {Beltr{\'a}n}, {Garc{\'\i}a},
  {Patino-Esteban}, {L{\'o}pez-Fern{\'a}ndez}, {G{\'o}mez-Molina}, {Diez},
  {Garc{\'\i}a-Carre{\~n}o}, {Malo}, {Amils}, {Serna}, {Albo}, {Hern{\'a}ndez},
  {Vaquero}, {Gonz{\'a}lez-Garc{\'\i}a}, {Barbas}, {L{\'o}pez-Fern{\'a}ndez},
  {Bujarrabal}, {G{\'o}mez-Garrido}, {Pardo}, {Santander-Garc{\'\i}a},
  {Tercero}, {Cernicharo}, \& {de Vicente}}]{TerceroF2020}
{Tercero}, F., {L{\'o}pez-P{\'e}rez}, J.~A., {Gallego}, J.~D., {et~al.}
  2020{\natexlab{b}}, arXiv e-prints, arXiv:2010.16224

\end{thebibliography}

\begin{appendix}
\section{Additional figures and tables}
\label{appen_figsTables}

\begin{table*}
\begin{center}
\caption{Observed lines of C$_4$H$_3$N isomers towards TMC-1 (CP).}
\label{table:fits}
\resizebox{.82\textwidth}{!}{
\begin{tabular}{llrrrlllr}
\hline\hline       
Transition  & \multicolumn{1}{c}{Rest Freq.} & \multicolumn{1}{c}{$E{up}$}  & $A_{ij}$  & \multicolumn{1}{c}{$S_{ij}$} & \multicolumn{1}{c}{$\int T_{\rm A}^* dv$} & \multicolumn{1}{c}{$V_{\rm LSR}$} & \multicolumn{1}{c}{$\Delta$v} & \multicolumn{1}{c}{$T_{\rm A}^*$}\\
$(J_{K_{\rm a},K_{\rm c}})_{\rm u}-(J_{K_{\rm a},K_{\rm c}})_{\rm l}$ & \multicolumn{1}{c}{(MHz)}      & \multicolumn{1}{c}{(K)}  & \multicolumn{1}{c}{(10$^{-6}$ s$^{-1}$)} &             
                              &  \multicolumn{1}{c}{(K km s$^{-1}$)}  & \multicolumn{1}{c}{(km s$^{-1}$)} & \multicolumn{1}{c}{(km s$^{-1}$)} & \multicolumn{1}{c}{(K)}      \\
\hline
\multicolumn{9}{c}{HCCCH$_2$CN, $a$-type, $\mu_{\rm a}=2.87$\,D}\\
\hline
$6_{1,6}-5_{1,5}$  &   31848.982(3) &   6.2 &  1.39 &  5.83 &  0.0093(10) & 5.83( 2) & 0.79( 6) & 0.0111( 8) \\
$6_{0,6}-5_{0,5}$  &   32722.702(3) &   5.5 &  1.55 &  5.99 &  0.0099( 5) & 5.67( 1) & 0.78( 3) & 0.0120( 4) \\
$6_{2,5}-5_{2,4}$  &   32876.187(3) &   8.8 &  1.40 &  5.33 &  0.0025( 6) & 5.51( 6) & 0.66(14) & 0.0036( 5) \\
$6_{2,4}-5_{2,3}$  &   33048.726(3) &   8.8 &  1.42 &  5.33 &  0.0016( 6) & 5.49(10) & 0.70(20) & 0.0021( 5) \\
$6_{1,5}-5_{1,4}$  &   33863.716(3) &   6.5 &  1.67 &  5.83 &  0.0072( 5) & 5.66( 2) & 0.79( 4) & 0.0085( 4) \\
$7_{1,7}-6_{1,6}$  &   37139.207(4) &   8.0 &  2.24 &  6.86 &  0.0081( 5) & 5.74( 2) & 0.69( 3) & 0.0110( 5) \\
$7_{0,7}-6_{0,6}$  &   38102.698(4) &   7.3 &  2.47 &  6.99 &  0.0110( 6) & 5.70( 1) & 0.74( 3) & 0.0140( 5) \\
$7_{2,6}-6_{2,5}$  &   38342.339(4) &  10.6 &  2.32 &  6.43 &  0.0016( 6) & 5.56( 9) & 0.57(18) & 0.0027( 7) \\
$7_{2,5}-6_{2,4}$  &   38616.702(4) &  10.7 &  2.37 &  6.43 &  0.0039( 5) & 5.64( 3) & 0.60( 5) & 0.0061( 6) \\    
$7_{1,6}-6_{1,5}$  &   39486.580(4) &   8.4 &  2.70 &  6.86 &  0.0075( 5) & 5.66( 2) & 0.60( 3) & 0.0117( 6) \\
$8_{1,8}-7_{1,7}$  &   42421.779(4) &  10.0 &  3.39 &  7.87 &  0.0077( 8) & 5.74( 2) & 0.59( 5) & 0.0122( 9) \\    
$8_{0,8}-7_{0,7}$  &   43450.742(4) &   9.4 &  3.70 &  7.99 &  0.0109(11) & 5.73( 2) & 0.73( 5) & 0.0140(10) \\
$8_{2,7}-7_{2,6}$  &   43802.419(4) &  12.7 &  3.55 &  7.50 &  0.0028( 8) & 5.69( 4) & 0.48(10) & 0.0056(10) \\
$8_{2,6}-7_{2,5}$  &   44210.195(4) &  12.8 &  3.66 &  7.50 &  0.0018(16) & 5.70( 5) & 0.35(31) & 0.0047(11) \\
$8_{1,7}-7_{1,6}$  &   45099.074(4) &  10.6 &  4.07 &  7.87 &  0.0125( 9) & 5.69( 1) & 0.61( 3) & 0.0192(10) \\
$9_{1,9}-8_{1,8}$  &   47696.032(5) &  12.3 &  4.86 &  8.89 &  0.0036(13) & 5.74( 6) & 0.50(13) & 0.0068(17) \\
$9_{0,9}-8_{0,8}$  &   48764.484(5) &  11.8 &  5.26 &  8.98 &  0.0078(16) & 5.76( 4) & 0.63(10) & 0.0117(14) \\
\hline
\multicolumn{9}{c}{HCCCH$_2$CN, $b$-type, $\mu_{\rm b}=2.19$\,D}\\
\hline
$3_{1,3}-2_{0,2}, F_{\rm u}-F_{\rm l}=3-2$  &  32519.775(3) &    2.3 & 0.49 &  1.79   & \rdelim\}{1.5}{*}[0.0031( 8)]  &  \multirow{2}{*}{5.59( 9)$^*$}   &  \multirow{2}{*}{1.06(20)} &  \multirow{2}{*}{0.0028( 5)}  \\ 
$3_{1,3}-2_{0,2}, F_{\rm u}-F_{\rm l}=2-1$  &  32519.815(3) &    2.3 & 0.46 &  1.21   &           &          & &   \\
$3_{1,3}-2_{0,2}, F_{\rm u}-F_{\rm l}=4-3$  &  32519.916(3) &    2.3 & 0.55 &  2.58   & 0.0013( 4) & 5.73( 7) & 0.58(22) & 0.0021( 4) \\
%
$9_{0,9}-8_{1,8},  F_{\rm u}-F_{\rm l}=8-7$  &  36933.586(4) &   11.8 & 0.74 &  4.44  & \rdelim\}{1.5}{*}[0.0034(17)]  &  \multirow{2}{*}{5.80(11)$^*$}   &  \multirow{2}{*}{1.16(31)} &  \multirow{2}{*}{0.0027( 6)}  \\
$9_{0,9}-8_{1,8}, F_{\rm u}-F_{\rm l}=10-9$  &  36933.621(4) &   11.8 & 0.75 &  5.57   &            &          & &   \\ 
$9_{0,9}-8_{1,8},  F_{\rm u}-F_{\rm l}=9-8$  &  36933.800(4) &   11.8 & 0.74 &  4.98   & 0.0020( 8) & 5.86(12) & 0.84(20) & 0.0023( 6) \\
%
$4_{1,4}-3_{0,3}, F_{\rm u}-F_{\rm l}=4-3$  &  37340.269(4) &    3.4 & 0.77 &  2.38   & 0.0025( 4) & 5.78( 7) & 0.82(13) & 0.0029( 6) \\
$4_{1,4}-3_{0,3}, F_{\rm u}-F_{\rm l}=3-2$  &  37340.455(4) &    3.4 & 0.75 &  1.81  & \rdelim\}{1.5}{*}[0.0030( 3)]  &  \multirow{2}{*}{5.78( 4)$^*$}   &  \multirow{2}{*}{0.60( 7)} &  \multirow{2}{*}{0.0047( 5)}  \\
$4_{1,4}-3_{0,3}, F_{\rm u}-F_{\rm l}=5-4$  &  37340.473(4) &    3.4 & 0.82 &  3.10  &            &          & &   \\ 
%
$5_{1,5}-4_{0,4}, F_{\rm u}-F_{\rm l}=5-4$  &  42010.978(4) &    4.6 & 1.12 &  2.96   & 0.0017( 5) & 5.85( 6) & 0.46(19) & 0.0035(10) \\
$5_{1,5}-4_{0,4}, F_{\rm u}-F_{\rm l}=6-5$  &  42011.211(4) &    4.6 & 1.16 &  3.65  & \rdelim\}{1.5}{*}[0.0028(13)]  &  \multirow{2}{*}{5.80( 6)$^*$}   &  \multirow{2}{*}{0.54(15)} &  \multirow{2}{*}{0.0050(10)}  \\ 
$5_{1,5}-4_{0,4}, F_{\rm u}-F_{\rm l}=4-3$  &  42011.215(4) &    4.6 & 1.10 &  2.40   &            &          & &   \\
%
$6_{1,6}-5_{0,5}, F_{\rm u}-F_{\rm l}=6-5$  &  46545.801(5) &    6.2 & 1.55 &  3.57   & 0.0013( 5) & 5.63(11) & 0.48(20) & 0.0026(12) \\
$6_{1,6}-5_{0,5}, F_{\rm u}-F_{\rm l}=7-6$  &  46546.045(5) &    6.2 & 1.59 &  4.24  & \rdelim\}{1.5}{*}[0.0041( 7)]  &  \multirow{2}{*}{5.69( 6)$^*$}   &  \multirow{2}{*}{0.54( 9)} &  \multirow{2}{*}{0.0071(11)}  \\ 
$6_{1,6}-5_{0,5}, F_{\rm u}-F_{\rm l}=5-4$  &  46546.057(5) &    6.2 & 1.54 &  3.00   &            &          & &   \\
%
\hline
\multicolumn{9}{c}{CH$_2$CCHCN, $\mu_{\rm a}=4.07$\,D}\\
\hline
$6_{1,5}-5_{1,4}$  &   31615.627(5) &   6.4 &  2.73 &  5.83 &  0.0247(12) & 5.64( 1) & 0.80( 3) & 0.0289( 9) \\
$7_{1,7}-6_{1,6}$  &   35379.044(5) &   7.9 &  3.90 &  6.86 &  0.0213( 6) & 5.68( 1) & 0.77( 2) & 0.0259( 5) \\
$7_{0,7}-6_{0,6}$  &   36064.688(5) &   6.9 &  4.22 &  7.00 &  0.0253( 4) & 5.64( 1) & 0.69( 1) & 0.0347( 4) \\
$7_{2,6}-6_{2,5}$  &   36140.273(5) &  11.4 &  3.90 &  6.43 &  0.0086( 8) & 5.61( 3) & 0.90( 6) & 0.0090( 6) \\
$7_{2,5}-6_{2,4}$  &   36222.501(5) &  11.4 &  3.93 &  6.43 &  0.0095( 6) & 5.63( 2) & 0.90( 4) & 0.0099( 5) \\
$7_{1,6}-6_{1,5}$  &   36878.547(5) &   8.2 &  4.42 &  6.86 &  0.0226( 5) & 5.63( 1) & 0.70( 1) & 0.0300( 5) \\
$8_{1,8}-7_{1,7}$  &   40425.712(6) &   9.9 &  5.90 &  7.87 &  0.0198( 8) & 5.70( 1) & 0.60( 2) & 0.0311( 8) \\
$8_{0,8}-7_{0,7}$  &   41187.082(6) &   8.9 &  6.34 &  8.00 &  0.0245( 6) & 5.66( 1) & 0.59( 1) & 0.0388( 7) \\
$8_{2,7}-7_{2,6}$  &   41297.656(6) &  13.4 &  5.99 &  7.50 &  0.0088( 7) & 5.67( 2) & 0.65( 4) & 0.0128( 6) \\
$8_{2,6}-7_{2,5}$  &   41420.713(6) &  13.4 &  6.04 &  7.50 &  0.0084( 7) & 5.64( 2) & 0.65( 4) & 0.0121( 6) \\
$8_{1,7}-7_{1,6}$  &   42138.451(6) &  10.2 &  6.68 &  7.87 &  0.0194( 9) & 5.63( 1) & 0.52( 2) & 0.0348(10) \\
$9_{1,9}-8_{1,8}$  &   45469.519(6) &  12.0 &  8.48 &  8.89 &  0.0176( 9) & 5.69( 1) & 0.56( 2) & 0.0293(11) \\
$9_{0,9}-8_{0,8}$  &   46297.882(6) &  11.1 &  9.06 &  9.00 &  0.0205(11) & 5.67( 1) & 0.55( 2) & 0.0352(13) \\
$9_{2,8}-8_{2,7}$  &   46452.840(6) &  15.6 &  8.70 &  8.56 &  0.0097(10) & 5.74( 2) & 0.66( 5) & 0.0138(10) \\
$9_{2,7}-8_{2,6}$  &   46628.069(6) &  15.7 &  8.80 &  8.56 &  0.0083( 9) & 5.64( 2) & 0.60( 4) & 0.0129(11) \\
$9_{1,8}-8_{1,7}$  &   47394.848(6) &  12.5 &  9.60 &  8.89 &  0.0152(13) & 5.63( 2) & 0.62( 4) & 0.0232(13) \\
\hline
\multicolumn{9}{c}{CH$_3$CCCN, $\mu_{\rm a}=4.75$\,D}\\
\hline
E  $8_2-7_2$  &    33050.3475(8) & 29.4 & 4.18&  7.50  &  0.0044( 9) &5.14( 9) &1.41(16)&  0.0029( 5)  \\
E  $8_1-7_1$  &    33051.3033(9) &  6.9 & 4.39&  7.88  &  0.0472( 8) &5.79( 1) &0.76( 1)&  0.0580( 6)  \\
A  $8_0-7_0$  &    33051.6219(9) &  7.1 & 4.46&  8.00  &  0.0485( 8) &5.80( 1) &0.74( 1)&  0.0616( 6)  \\
E  $9_2-8_2$  &    37181.5838(9) & 31.2 & 6.08&  8.56  &  0.0013( 7) &5.77(10) &0.64(24)&  0.0020( 7)  \\
E  $9_1-8_1$  &    37182.659(1)  &  8.7 & 6.32&  8.89  &  0.0421( 8) &5.79( 1) &0.68( 1)&  0.0585( 7)  \\
A  $9_0-8_0$  &    37183.017(1)  &  8.9 & 6.40&  9.00  &  0.0455( 8) &5.79( 1) &0.68( 1)&  0.0632( 7)  \\
E $10_2-9_2$  &    41312.799(1)  & 33.2 & 8.46&  9.60  &  0.0025(11) &5.42( 8) &0.55(15)&  0.0031(10)  \\
E $10_1-9_1$  &    41313.994(1)  & 10.7 & 8.73&  9.90  &  0.0372(64) &5.81( 4) &0.58(10)&  0.0604( 9)  \\
A $10_0-9_0$  &    41314.393(1)  & 10.9 & 8.81& 10.0   &  0.0412(51) &5.83( 3) &0.57( 7)&  0.0674( 9)  \\
E $11_2-10_2$ &    45443.993(1)  & 35.4 & 11.4& 10.6   &  ...   &...      &...     & $\leq$0.0050(10)   \\
E $11_1-10_1$ &    45445.307(1)  & 12.9 & 11.7& 10.9   &  0.0294(42) &5.80( 3) &0.57( 8)&  0.0483(12)  \\
A $11_0-10_0$ &    45445.745(1)  & 13.1 & 11.8& 11.0   &  0.0321(42) &5.82( 3) &0.62( 8)&  0.0487(12)  \\
E $12_2-11_2$ &    49575.162(1)  & 37.7 & 14.9& 11.7   &  ...   &...      &...     & $\leq$0.0060(20)  \\
E $12_1-11_1$ &    49576.596(1)  & 15.3 & 15.2& 11.9   &  0.0242(52) &5.75( 5) &0.69(14)&  0.0328(26)  \\
A $12_0-11_0$ &    49577.073(1)  & 15.5 & 15.4& 12.0   &  0.0269(38) &5.81( 4) &0.64( 8)&  0.0393(26)  \\
\hline
\end{tabular}
}
\end{center}
\textbf{Notes.} 
$^*$ LSR velocity corresponds to the strongest hyperfine transition. 
Numbers in parentheses indicate the uncertainty in units of the last significant digits. 
For the observational parameters we adopted the uncertainty of the Gaussian fit provided by \texttt{GILDAS}. 
\textbf{HCCCH$_2$CN}: Spectroscopic line parameters were obtained using \texttt{MADEX}
by fitting the rotational lines reported by \citet{Demaison1985} and \citet{McNaughton1988}.
Dipole moments are from \citet{McNaughton1988}. 
\textbf{CH$_2$CCHCN}: Spectroscopic line parameters were obtained using \texttt{MADEX}
by fitting the rotational lines reported by \citet{Bouchy1973} and \citet{Schwahn1986}.
Dipole moment is from \citet{Bouchy1973}.
\textbf{CH$_3$CCCN}: Spectroscopic line parameters were obtained using \texttt{MADEX}
by fitting the rotational lines reported by \citet{Moises1982} and \citet{Bester1983}.
Rotation constants $A$ and $D_{\rm k}$ have been assumed to be the same as those of CH$_3$CN.
Some additional data have been taken from the CDMS (\texttt{https://cdms.astro.uni-koeln.de/}).
Dipole moment is from \citet{Bester1984}. Note that the E species is 7.8\,K above the A species, and energies for the E species are referred to the lowest energy level (1,1).
\end{table*}

\begin{figure*}
\centering
\includegraphics[width=0.95\textwidth]{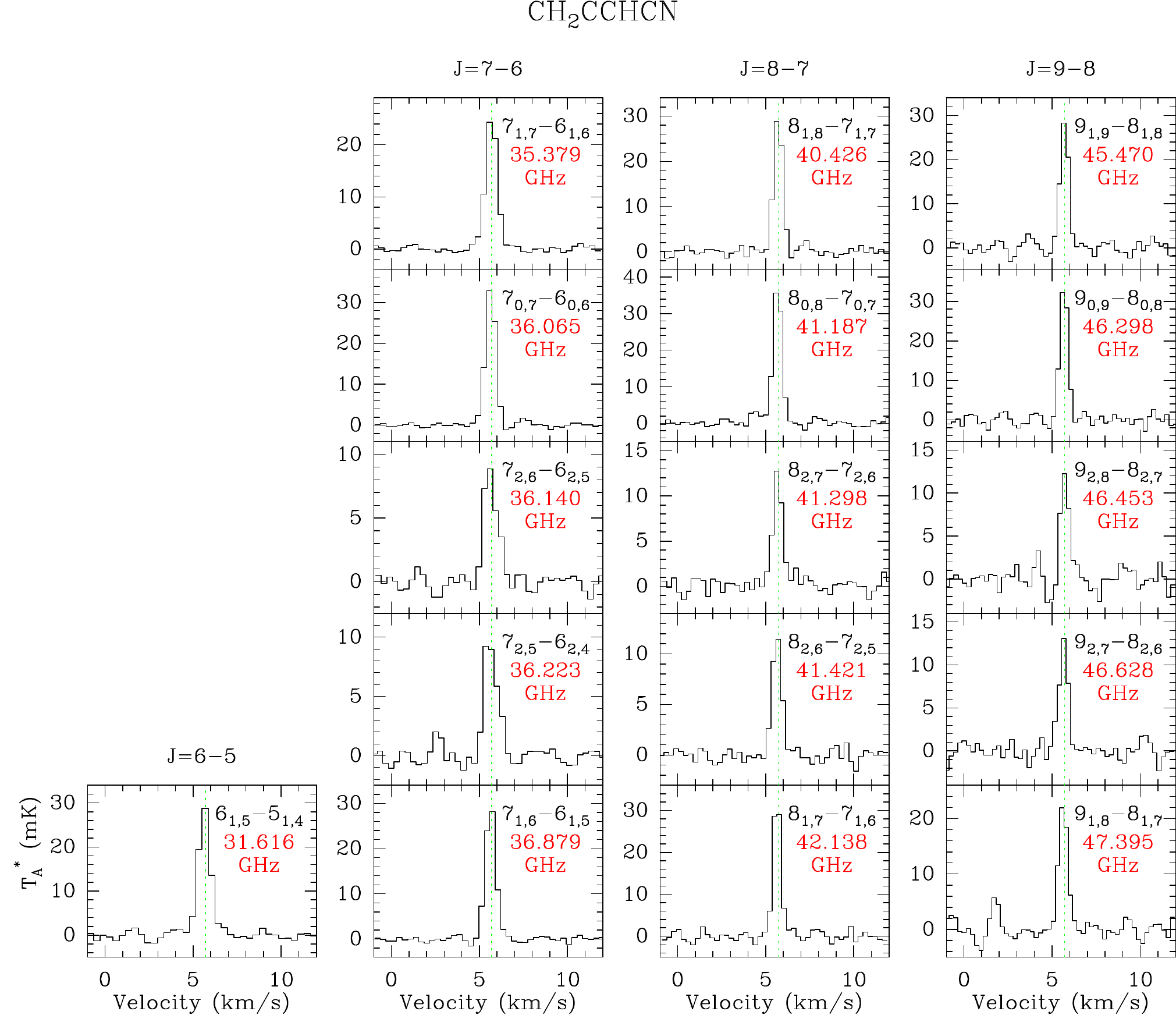}
\caption{Observed lines of CH$_2$CCHCN toward TMC-1 (CP). The vertical dashed green line marks a radial velocity 
of 5.7\,km\,s$^{-1}$.}
\label{fig:lines-CH2CCHCN}
\end{figure*}

\begin{figure*}
\centering
\includegraphics[width=0.95\textwidth]{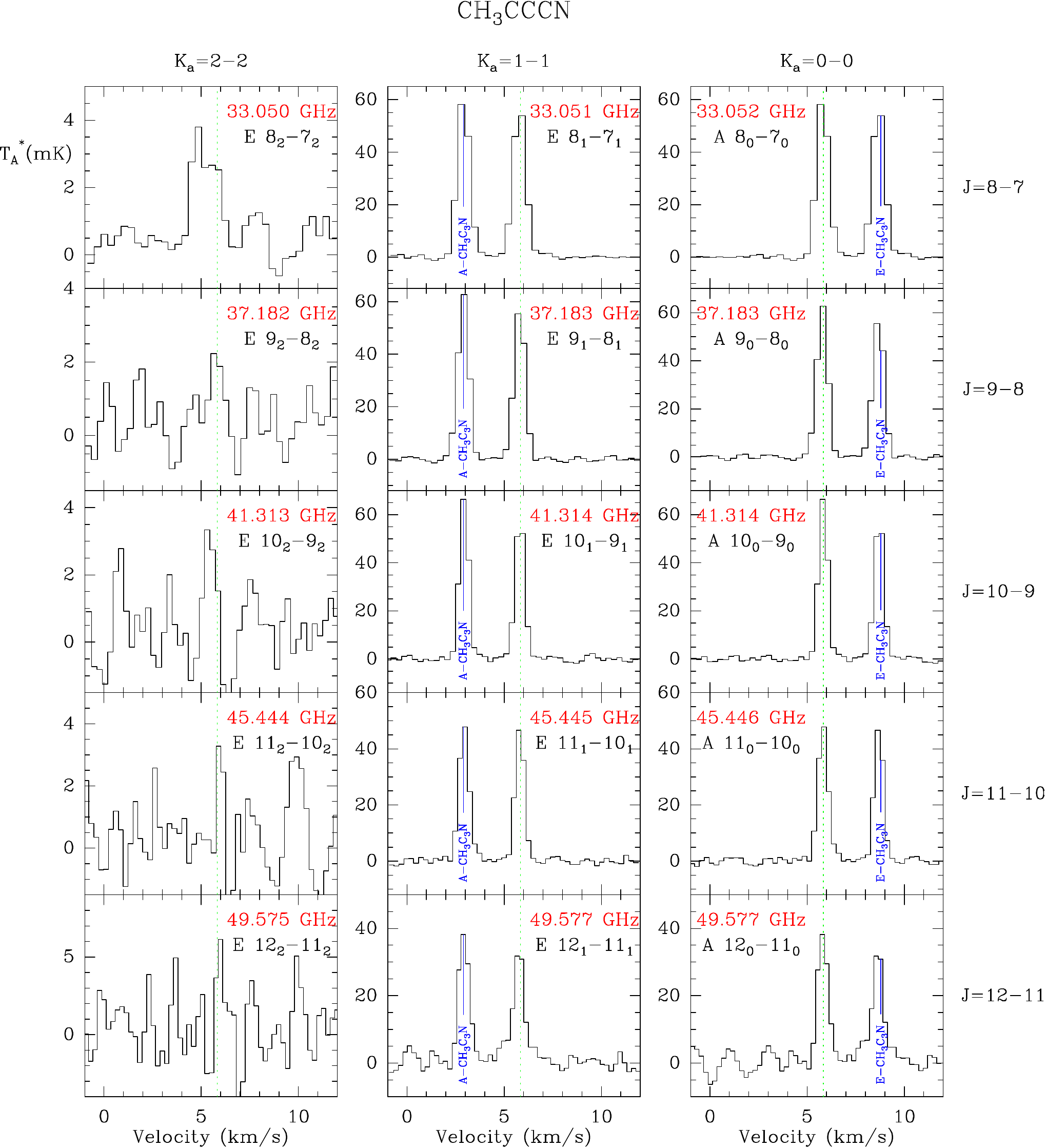}
\caption{Observed lines from CH$_3$CCCN towards TMC-1 (CP). Dashed green line marks a radial velocity 
of 5.8\,km\,s$^{-1}$. }
\label{fig:lines-CH3CCCN}
\end{figure*}

\begin{figure*}
\centering
\includegraphics[width=0.70\columnwidth]{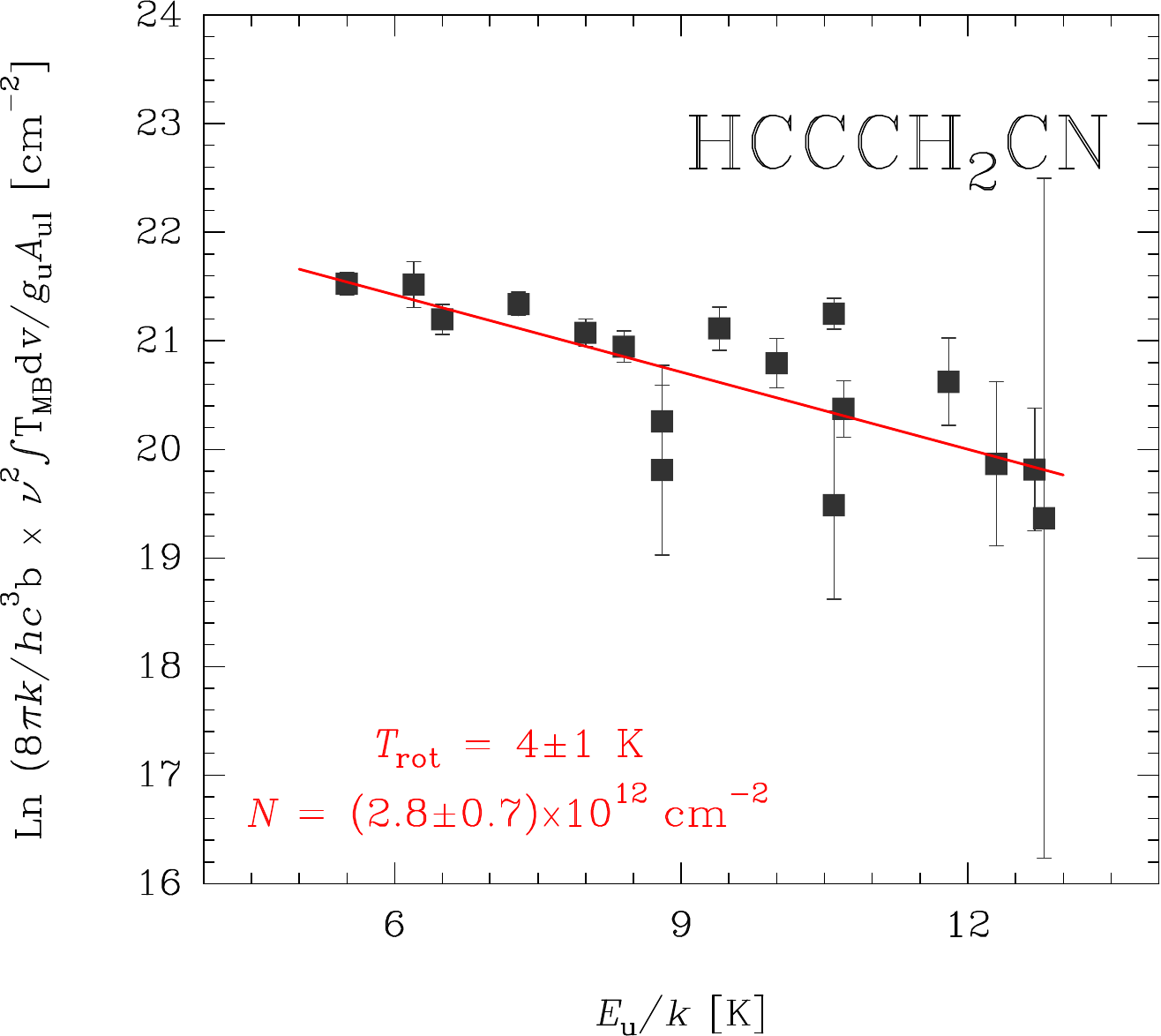} \hspace{0.7cm} \includegraphics[width=0.70\columnwidth]{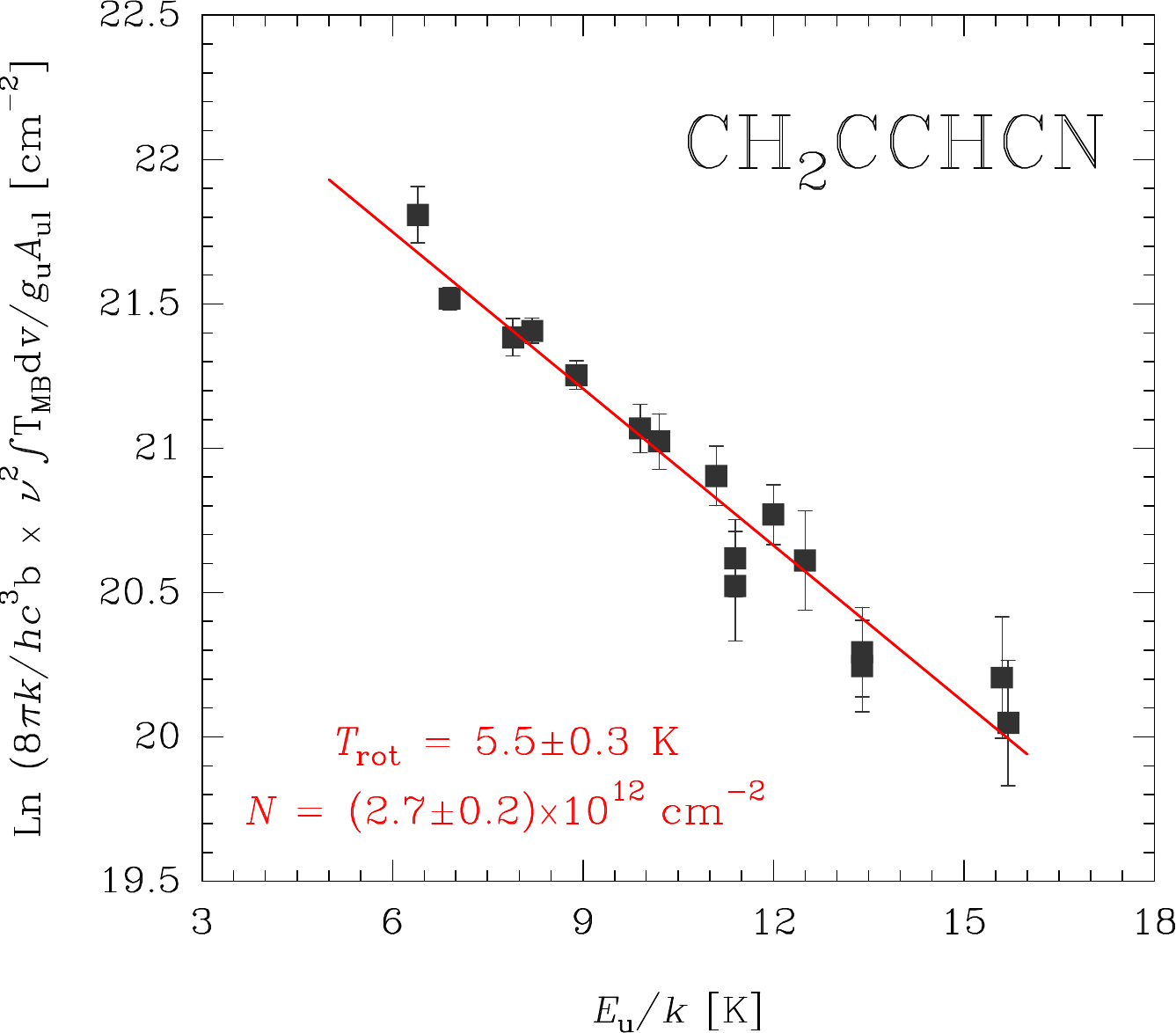}\\\includegraphics[width=0.70\columnwidth]{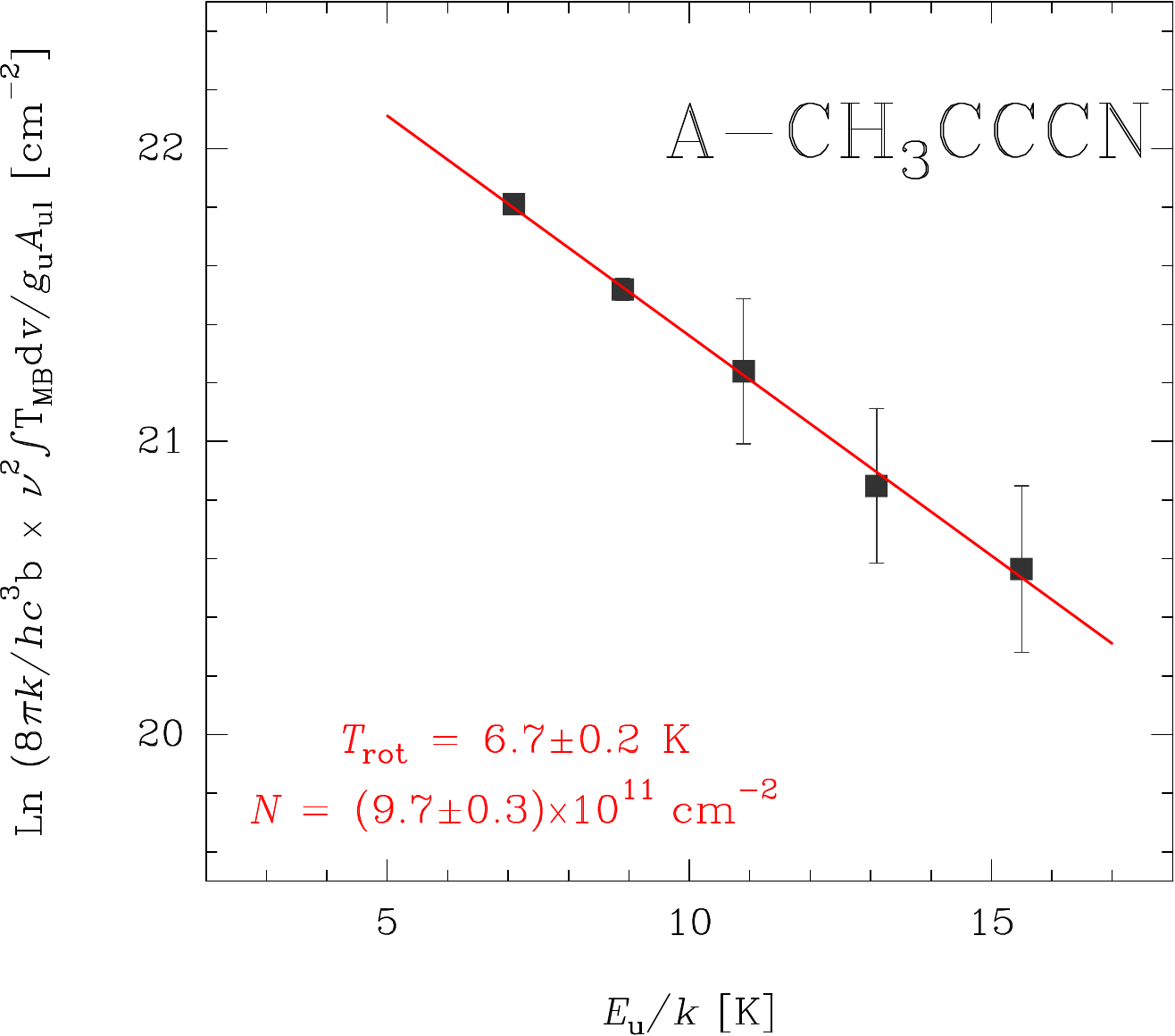}  \hspace{0.7cm} \includegraphics[width=0.70\columnwidth]{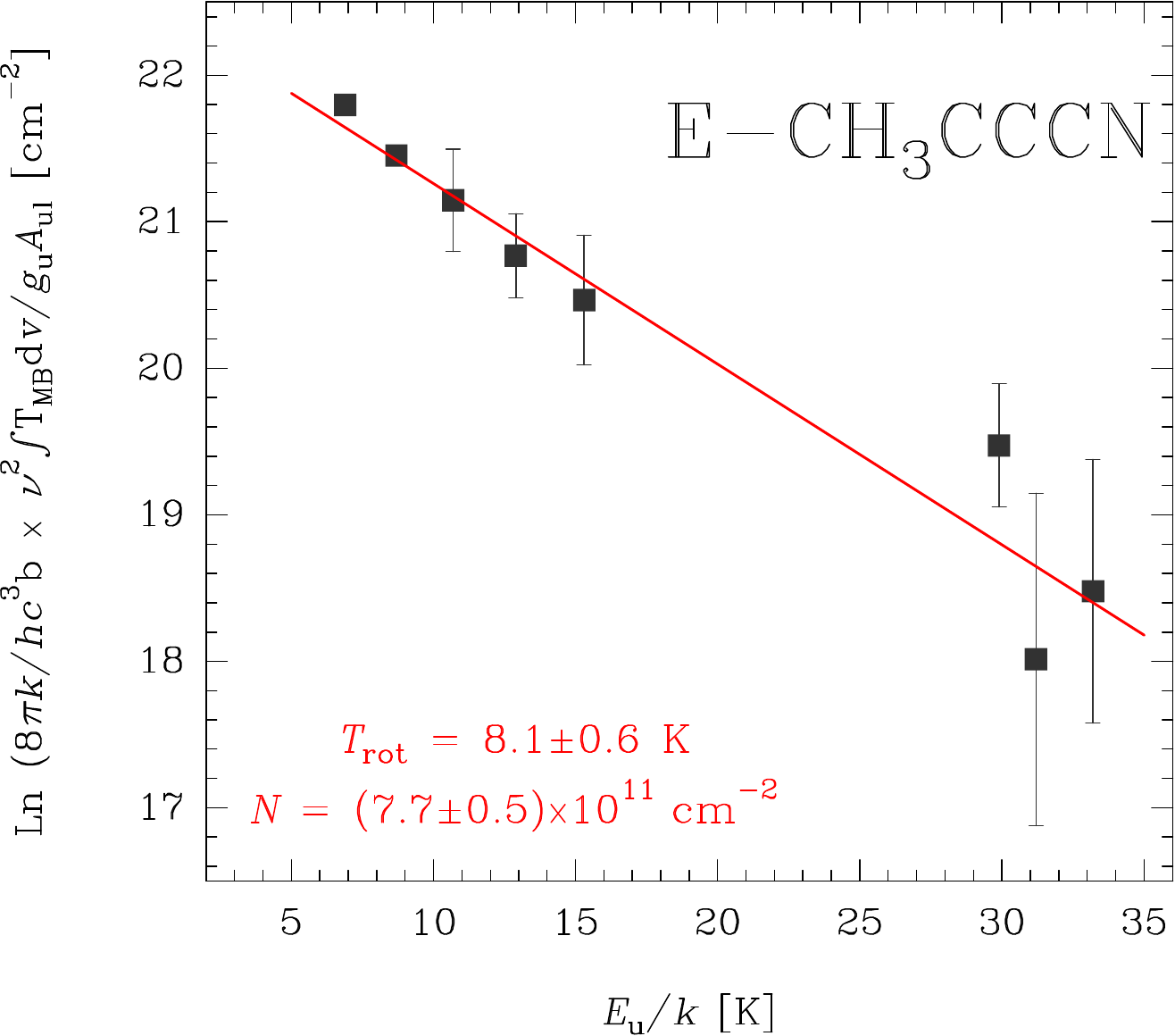}
\caption{Rotational diagrams of the C$_4$H$_3$N isomers towards TMC-1 (CP). Derived values of the rotational temperature, $T_{\rm rot}$, column density, $N$, and their respective uncertainties are indicated for each molecule.}\label{fig_DR}
\end{figure*}

\begin{figure}
\includegraphics[width=\columnwidth]{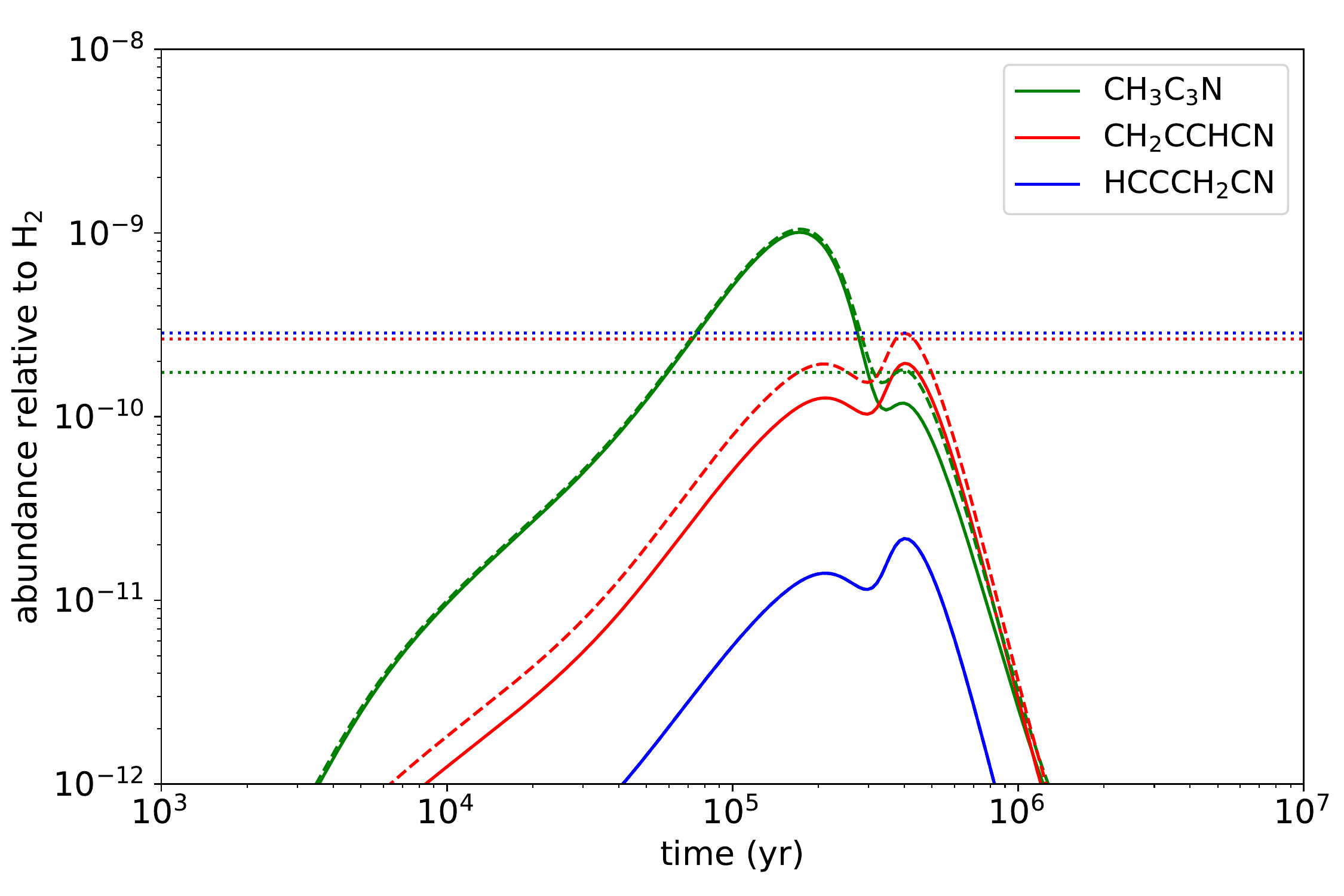}
\caption{Calculated fractional abundances of the three C$_4$H$_3$N isomers as a function of 
time. Solid and dashed lines correspond to two models in which we use branching ratios 
for the CN + CH$_3$CCH reaction from \cite{Abeysekera2015} and from \cite{Balucani2000}, 
respectively (see text). The abundances observed in TMC-1 for the three C$_4$H$_3$N isomers 
(from Table~\ref{table_cd} adopting a H$_2$ column density of 10$^{22}$ cm$^{-2}$; 
\citealt{Cernicharo1987}) are shown as horizontal dotted lines.}
\label{fig:abun}
\end{figure}

\begin{figure}
\includegraphics[width=6.5cm]{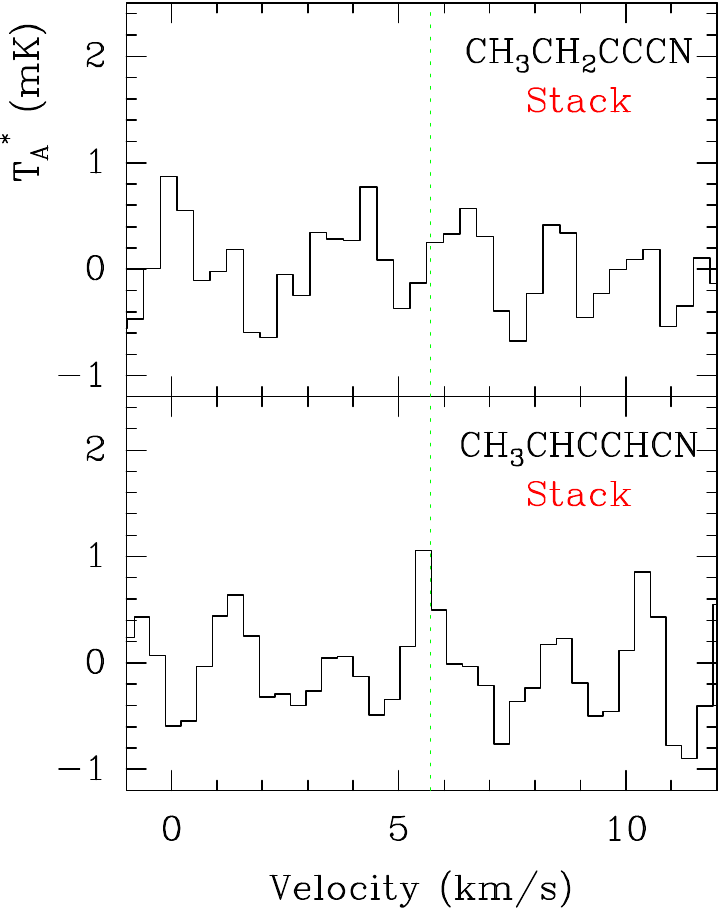}
\caption{Stacked spectra of CH$_3$CH$_2$CCCN and CH$_3$CHCCHCN toward TMC-1.}
\label{fig_stack}
\end{figure}

\end{appendix}

\end{document}